\documentclass{article}

\def\book#1{{\it #1,\/}}
\def\journal#1{{\it #1\/}}
\def\art#1{\edef\test{#1} \edef\control{}
      \ifx\test\control  {\bf {?!}}\message{warning?!}\else \\ #1,\fi}
\def\pub#1{#1}
\def\Ed#1{Ed.\ #1,}
\def\Eds#1{Eds.\ #1,}

\def\journalfont{\it} 
  \def\jou#1{{\journalfont #1\ }}

\def\joudef#1#2{\def #1{\jou{\ignorespaces #2}}}

\def\vol#1{{\bf #1}}

\joudef{\AAA}{  Astron.\ Astrophys.}
\joudef{\AIP}{  Adv.\ Phys.}
\joudef{\AJP}{  Amer.\ J.\ Phys.}
\joudef{\AM}{   Ann.\ Math.}
\joudef{\AP}{   Ann.\ Phys.\ (N.Y.)}
\joudef{\AOP}{  Ann.\ Phys.\ (N.Y.)}
\joudef{\APJ}{  Astrophys.\ J.}
\joudef{\APP}{  Acta Phys.\ Polonica.}
\joudef{\CJP}{  Can.\ J.\ Phys.}
\joudef{\CMP}{  Commun.\ Math.\ Phys.}
\joudef{\CQG}{  Class.\ Quantum Grav.}
\joudef{\GRG}{  Gen.\ Relativ.\ Grav.}
\joudef{\IJTP}{ Int.\  J.\ Theor.\ Phys.}
\joudef{\IJMP}{ Int.\  J.\ Mod.\ Phys.}
\joudef{\JKPS}{ J.\ Korean.\ Phys.\ Soc.}
\joudef{\JMP}{  J.\ Math.\ Phys.}
\joudef{\JPAMG}{J.\ Phys.\ A: Math.\ Gen.}
\joudef{\MNRAS}{Mon.\ Not.\ R.\ Ast.\ Soc.}
\joudef{\NAT}{  Nature}
\joudef{\NCIM}{ Nuovo Cim.}
\joudef{\NUCP}{ Nuc.\ Phys.}
\joudef{\NCB}{  Il Nuovo Cimento B}
\joudef{\PL}{   Phys.\ Lett.}
\joudef{\PR}{   Phys.\ Rev.}
\joudef{\PREP}{ Phys.\ Rep.}
\joudef{\PRL}{  Phys.\ Rev.\ Lett.}
\joudef{\PTP}{  Prog.\ Theor.\ Phys.}
\joudef\RMP{    Rev.\ Mod.\ Phys.}
\joudef\SPJ{    Sov.\ Phys.\ JETP}

\joudef\GRQC{electronic preprint: {\rm gr-qc/}}
\joudef\ASTROPH{electronic preprint: {\rm astro-ph/}}

\begin{document}

\begin{center}
{\Huge
{\bf
A List of References on \\
Spacetime Splitting \\
\bigskip
and Gravitoelectromagnetism
}}
\\[24pt]
Donato Bini\dag ${}^\ast$ and Robert T. Jantzen$\S$ ${}^\ast$\\
\end{center}

\begin{flushleft}
\dag\ {\it
Istituto per Applicazioni della Matematica C.N.R.,
I--80131 Napoli, Italy} \\
${}^\ast$ {\it
 I.C.R.A., International Center for Relativistic Astrophysics, University of Rome, I--00185 Rome, Italy}\\
$\S$ {\it
  Department of Mathematical Sciences, Villanova University, 
  Villanova, PA 19085, USA}
\end{flushleft}

To be published in the 
Proceedings of the Encuentros Relativistas Espa\~noles, September, 2000
 
(http://hades.eis.uva.es/EREs2000).

We invite corrections and additions to be sent to {\tt robert.jantzen@villanova.edu}. 

\begin{enumerate}

\item [A]

\item[ ] Abramowicz, M.A., 
  see also Kristiansson, Sonego.

\item[ ] Abramowicz, M.A., 1990,
  \art{Centrifugal Force: A Few Surprises}
  \MNRAS \vol{245}, 733.

\item[ ] Abramowicz, M.A., 1992,
  \art{Relativity of Inwards and Outwards}
  \MNRAS \vol{256}, 710.

\item[ ] Abramowicz, M.A.,  1993, 
  \art{Black Holes and the Centrifugal Force Paradox}
  \journal{Sci.\ Am.} \vol{268}, 26. 

\item[ ] Abramowicz, M.A. and Bi\v c\'ak, J., 1991,
  \art{The Interplay between Relativistic Gravitational,
       Centrifugal and Electric Forces: a Simple Example}
  \GRG \vol{23}, 941.

\item[ ] Abramowicz, M.A., Carter, B. and Lasota, J.P., 1988,
  \art{Optical Reference Geometry for Stationary and Static Dynamics}
  \GRG \vol{20}, 1173.

\item[ ] Abramowicz, M.A. and Lasota, J.P., 1974,
   \art{A Note on a Paradoxical Property of the Schwarzschild Solution}
   \APP \vol{B5}, 327.

\item[ ] Abramowicz, M.A.  and Lasota, J.P.,  1986, 
  \art{On Traveling Round Without Feeling It and Uncurving Curves}
  \AJP \vol{54}, 936. 

\item[ ] Abramowicz, M.A. and Lasota, J.P., 1997,
   \art{A Brief Story of a Straight Circle}
   \CQG \vol{14}, A23.

\item[ ] Abramowicz, M.A., Miller, J.C., and Stuchl\'\i k, Z., 1993,
  \art{Concept of Radius of Gyration in General Relativity}
  \PR \vol{D47}, 1440.

\item[ ] Abramowicz, M.A., Nurowski, P., Wex, N., 1993,
   \art{Covariant Definition of Inertial Forces}
   \CQG \vol{10}, L183.

\item[ ] Abramowicz, M.A., Nurowski, P., Wex, N., 1995,
   \art{Optical reference geometry for stationary and axially symmetric 
        spacetimes} 
   \CQG \vol{12}, 1467.

\item[ ] Abramowicz, M.A. and Prasanna, A.R., 1990,
  \art{Centrifugal-force Reversal Near a Schwarzschild Black Hole}
  \MNRAS \vol{245}, 720.

\item[] Abramowicz, M.A., 1990,
  \art{Unexpected Properties of the Centrifugal Force}
in 1989 Summer School in High Energy Physics and Cosmology,
Pati, J.C., Randjbar-Daemi, S., Sezgin, E., Shafi, Q. (editors),
World Scientific, Singapore, 586.

\item[] Abramowicz, M.A., 1992,
  \art{Relativity of Inwards and Outwards}
\MNRAS \vol{256}, 710.

\item[] Abramowicz, M.A., 1993,
  \art{Inertial Forces in General Relativity}
in \book{The Renaissance of General Relativity}
Ellis, G.F.R., Lanza, A., and Miller, J.C. (editors),
Cambridge University Press, Cambridge.

\item[ ] Aguirregabiria, J.M.,  Chamorro, A.,  Nayak, K.R.,
  Suinaga, J.  and Vishveshwara, C.V.,  1996, 
  \art{Equilibrium of a Charged Test Particle in the Kerr-Newman 
       Spacetime: Force Analysis}
  \CQG \vol{13}, 2179. 

\item[ ] Allen, B.,  1990, 
  \art{Reversing Centrifugal Forces}
  \journal{Nature} \vol{347}, 615. 

\item[ ] Anderson, R.,  Bilger, H.R. and  Stedman, G.E., 1994,
  \art{``Sagnac Effect: A Century of Earth-Rotated Interferometers}
  \journal{R.\ Am.\ J.\ Phys.} \vol{62}, 975.

\item[ ] Arms, J.A., 1979, 
  \art{Linearization stability of gravitational and gauge fields}  
  \JMP \vol{20}, 443.

\item[ ] Arms, J.A., Marsden, J.E., and Moncrief, V., 1982, 
  \art{The Structure of the Space of Solutions of Einstein's Equations II:
  Several Killing Fields and the Einstein-Yang-Mills Equations}
  \AOP \vol{144}, 81.

\item[ ] Arnowit, R., Deser, S. and Misner, C.W., 1962, 
  \art{The Dynamics of General Relativity}
  in \book{Gravitation: An Introduction to Current Research} 
  \Ed{L. Witten}
  \pub(Wiley, New York).

\item[ ] Ashby, N., and Shahid-Saless, B., 1990,
  \art{Geodetic Precession or Dragging of Inertial Frames?}
  \PR \vol{D42}, 1118.

\item[ ] Ashtekar, A.,  and Magnon, A., 1975, 
  \art{The Sagnac Effect in General Relativity} 
  \JMP \vol{16}, 341.

\item [B]
\item[ ] Bao, D., Marsden, J. and Walton, R., 1985, 
  \art{The Hamiltonian Structure of General Relativistic Perfect Fluids}   
  \CMP \vol{99}, 319.

\item[ ] Bardeen, J.M., 1970, 
  \art{Variational Principle for Rotating Stars in General Relativity}   
  \APJ \vol{162}, 71. 

\item[ ] Bardeen, J.M., 1980, 
  \art{Gauge-invariant Cosmological Perturbations}
  \PR \vol{D22}, 1882.

\item[ ] Bardeen, J.M., Press, W.H.  and Teukolsky, S.A., 1972,
  \APJ \vol{170}, 347.

\item[ ] Barnes, A., and Rowlingson, R.R., 1989,
  \art{Irrotational Perfect Fluids With a Purely Electric Weyl Tensor}
  \CQG \vol{6}, 949.

\item[ ] Barrab\`es, C., Boisseau, B., and Israel, W., 1995,
  \art{Orbits, Forces, and Accretion Dynamics Near Spinning Black Holes}
  \MNRAS \vol{276}, 432.

\item[ ] Baza\'nski, S.L., 1997,
  \art{Is the Geodesic Hypothesis in General Relativity 
        Falsifiable?}
  in \book{Mathematics of Gravitation Part II, Gravitational Wave 
           Detection, Banach Center Publications, Volume 41}
  \pub(Polish Academy of Sciences, Warsaw).

\item[ ] Belasco, E.P. and Ohanian, H.C., 1969, 
  \art{Initial Conditions in General Relativity: Lapse and Shift Formulation} 
  \JMP \vol{10}, 1503.

\item[ ] Benvenuti, P., 1960, 
  \art{Formulazione relativa delle equazioni dell'elettromagnetismo 
    in relativit\`a generale}
  \journal{Ann.\ Scuola Normale Sup.\ Pisa, Ser.III} \vol{14}, 171.

\item[ ] Bertschinger, E., and Hamilton, A.J.S., 1994,
  \art{Lagrangian Evolution of the Weyl Tensor}
 \APJ \vol{435}, 1.

\item[ ] Bini, D., Carini, P., and Jantzen, R.T., 1992,
  \art{Applications of Gravitoelectromagnetism to 
       Rotating Spacetimes}
  in \book{Proceedings of the Third Italian-Korean 
        Astrophysics Meeting}
  \Eds{S. Kim, H. Lee, K.T. Kim}
  \JKPS \vol{25}, S190.

\item[ ] Bini, D., Carini, P., and Jantzen, R.T., 1995,
  \art{Relative Observer Kinematics in General Relativity}
  \CQG \vol{12}, 2549.

\item[ ] Bini, D., Carini, P., and Jantzen, R.T., 1996,
  \art{Gravitoelectromagnetism: Further Applications}
  in \book{Proceedings of the Seventh Marcel Grossmann Meeting 
        on General Relativity (1994)}
  \Eds{R.T. Jantzen and G.M. Keiser} 
  \pub(World Scientific, Singapore), 519.

\item[ ] Bini, D., Carini, P., and Jantzen, R.T., 1997,
  \art{The Intrinsic Derivative and Centrifugal Forces. I: 
       Theoretical Foundations}
  \IJMP \vol{D6}, 1.

\item[ ] Bini, D., Carini, P., and Jantzen, R.T., 1997,
  \art{The Intrinsic Derivative and Centrifugal Forces. II: 
       Applications to Some Familiar Stationary Axisymmetric 
       Spacetimes}
  \IJMP \vol{D6}, 143.

\item[ ] Bini, D., Carini, P., Jantzen, R.T., and D. Wilkins, 1994,
  \art{Thomas Precession in Post-Newtonian 
       Gravitoelectromagnetism}
  \PR  \vol{D49}, 2820.

\item [] Bini, D., Jantzen, R.T., Merloni, A., 1998,
  \art{Adapted Frames for Spacetime Splitting with an 
       Additional Observer Family} 
\NCB \vol{113}, 611.

\item []  Bini, D., Jantzen, R.T., Merloni, A., 1999,
  \art{Geometric Interpretation of the Frenet-Serret Frame 
        Description of Circular Orbits in Stationary Axisymmetric                 Spacetimes} 
  \CQG \vol{16}, 1333.

\item []  Bini, D., de Felice, F., Jantzen, R.T., 1999,
  \art{Absolute and Relative Frenet-Serret Frames and 
       Fermi-Walker Transport} 
  \CQG \vol{16}, 2105.

\item []  Bini, D., Gemelli, G., Ruffini, R., 2000
\art{Spinning Test Particles in General Relativity: 
Nongeodesic Motion in the Reissner-Nordstr\"om Spacetime} 
\PR \vol{D61}, 064013-1.

\item [] Bini, D., Jantzen, R.T., 2000,
\art{Circular Orbits in Kerr Spacetime: Equatorial Plane Embedding Diagrams}
\CQG \vol{ 17}, 1. 

\item[ ] Blanchet, L. and Damour, T., 1989 
  \art{Post-Newtonian Generation of Gravitational Waves}
  \journal{Ann.\ Inst.\ Henri Poincar\'e} \vol{50}, 377.

\item[ ] Blanchet, L., Damour, T. and Sch\"afer, G., 1990,
  \art{Post-Newtonian Hydrodynamics and Post-Newtonian Gravitational
    Wave Generation for Numerical Relativity}
  \MNRAS \vol{242}, 289.

\item[ ] Boersma, S., and Dray, T., 1995,
  \art{Parametric Manifolds I: Extrinsic Approach}
  \JMP \vol{36}, 1378.

\item[ ] Boersma, S., and Dray, T., 1995,
  \art{Parametric Manifolds II: Intrinsic Approach}
  \JMP \vol{36}, 1394.

\item[ ] Boersma, S., and Dray, T., 1994,
  \art{Slicing, Threading and Parametric Manifolds}
  \GRG \vol{27}, 319.

\item[ ] Bonnor, W.B., 1992,
\GRG \vol{24}, 551.

\item[ ] Bonnor, W.B. and Steadman, B.R., 1999,
 \art{The Gravitomagnetic Clock Effect}
  \CQG \vol{16}, 1853.

\item [ ] Bradley, M.   Fodor, G.,  Gergely, L.A., Marklund, M. and Perj\'es, Z., 1999, 
  \CQG \vol{16}, 1667. 

\item[ ] Boyer, R.H., and  Lindquist, R.W., 1967, 
  \art{Maximal Analytic Extension of the Kerr Metric}  
  \JMP \vol{8}, 265.

\item[ ] Braginsky, V.B., Caves, C.M., and Thorne, K.S., 1977, 
  \art{Laboratory Experiments to Test Relativistic Gravity}
  \PR \vol{15}, 2047.

\item[ ] Braginsky, V.B., Polnarev, A.G., and Thorne, K.S., 1984,
  \art{Foucault Pendulum at the South Pole: Proposal For an Experiment to 
    Detect the Earth's General Relativistic Gravitomagnetic Field}
  \PRL \vol{53}, 863.

\item[ ] Bruni, M., Dunsby, P.K.S., and Ellis, G.F.R., 1992,
  \art{Cosmological Perturbations and the Meaning of Gauge-invariant Variables}
  \APJ \vol{395}, 34.

\item[ ] Bruni, M., Matarrese, S., and Pantano, O., 1994,
  \art{Dynamics of silent universes with and without lambda}
SISSA-171-94-A
{\it Talk given at 11th Italian Congress of General Relativity and Gravitation, Trieste, Italy, 26-30 Sep 1994}.

\item[ ] Bruni, M., Matarrese, S., and Pantano, O., 1995,
  \art{A Local View of the Observable Universe}
  \PRL \vol{74}, 1916.

\item[ ] Bunchaft, F., and Carneiro, S., 1998,
  \art{The static spacetime relative acceleration for the 
       general free fall and its possible experimental test}
  \CQG \vol{15}, 1557.

\item[ ] Burinskii, Kerr, R.P., and Perj\'es, Z., 1995,
  \art{Nonstationary Kerr Congruences}
\GRQC 9501012.

\item [C]
\item[ ] Calv\~ao, M.O., Soares, I.D. and Tiomno, J., 1990,
  \art{Geodesics in G\"odel-type Spacetimes}   
  \GRG \vol{22}, 683.

\item[ ] Carini, P., Bini, D., and Jantzen, R.T., 1992,
  \art{Gravitoelectromagnetism: Relativity of Splitting 
       Formalisms}
  in \book{Proceedings of the Third Italian-Korean 
        Astrophysics Meeting}
  \Eds{S. Kim, H. Lee, K.T. Kim}
  \JKPS \vol{25}, S233.

\item[ ] Carini, P., Bini, D., and Jantzen, R.T., 1996,
  \art{Gravitoelectromagnetism and Inertial Forces in 
       General Relativity}
  in \book{Proceedings of the Seventh Marcel Grossmann Meeting 
        on General Relativity (1994)}
  \Eds{R.T. Jantzen and G.M. Keiser} 
  \pub(World Scientific, Singapore), 522.

\item[ ] Carini, P. and Jantzen, R.T., 1993,
  \art{Gravitoelectromagnetism and the Single Gyro}
  in \book{Proceedings of the First William Fairbank Meeting 
        on Relativistic Experiments in Space (September 1990)}
  \Eds{R. Ruffini and M. Demianski}
  \pub(World Scientific Press, Singapore), 135.

\item[ ]
Carter, B., 1968, 
\art{Global Structure of the Kerr Family of Gravitational
fields}
\PR \vol{174}, 1559.

\item[ ] Carter, B., 1968, 
  \art{Hamilton-Jacobi and Schrodinger Separable Solutions of Einstein's 
     Equations}
  \CMP \vol{10}, 280.

\item[ ] Carter, B., 1969, 
  \art{Killing Horizons and Orthogonally Transitive Groups in Space-Time}  
  \JMP \vol{10}, 70.

\item[ ] Carter, B., 1973, 
  \art{Black Hole Equilibrium States}
  in \book{Black Holes}
  \Eds{C. DeWitt and B.S. DeWitt}
  \pub(Gordon and Breach, New York), 57.

\item[ ] Carter, B., McLenaghan, R.G., 1982, 
  \art{Generalized Master Equations for Wave Equation Separation in a Kerr 
    or Kerr-Newmann Black Hole Background}
  in \book{Proceedings of the Second Marcel Grossmann Meeting on General 
    Relativity}
  \Ed{R. Ruffini}
  \pub(North Holland, Amsterdam), p. 575.

\item[ ] Cattaneo, C., 1958, 
  \art{General Relativity: Relative Standard Mass,
    Momentum, Energy and Gravitational Field in a General System of Reference}
  \NCIM \vol{10}, 318.

\item[ ] Cattaneo, C., 1959, 
  \art{On the Energy Equation for a Gravitating
    Test Particle}   \NCIM \vol{11}, 733.

\item[ ] Cattaneo, C., 1959, 
  \art{Proiezioni naturali e derivazione traversa
    in una variet\`a riemanniana a metrica iperbolica normale}
  \journal{Ann.\ Mat.\ Pura ed Appl.} \vol{48}, 361.

\item[ ] Cattaneo, C., 1959, 
  \art{D\'erivation transverse et grandeurs relatives
    en relativit\'e g\'en\'erale}  
  \journal{Compt.\ Rend.\ Acad.\ Sci.} \vol{248}, 197.

\item[ ] Cattaneo, C., 1961, 
  \book{Calcolo Differenziale Assoluto su una Variet\'a Riemanniana}
  \pub(Libreria E.V. Veschi, Rome).

\item[ ] Cattaneo-Gasperini, I., 1961, 
  \art{Projections naturelles des tenseurs de courbure d'une vari\'et\'e 
     $V_{n+1}$ \`a m\'etrique hyperbolique normale}
  \journal{Compt.\ Rend.\ Acad.\ Sci.} \vol{252}, 3722.

\item[ ] Cattaneo-Gasperini, I., 1963, 
  \art{Proiezioni dei tensori di curvatura di una variet\'a riemanniana 
     a metrica iperbolica normale}
  \journal{Rend.\ di Mat.} \vol{22}, 127.

\item[ ] Chakrabarti, S.K.,
  see also Prasanna.

\item[ ] Chakrabarti, S.K.  and Prasanna, A.R., 1990,
  \art{Classical Forces in the Kerr Geometry}  
  \journal{J.\ Astrophys.\ Astro.} \vol{11}, 29.

\item[ ] Chakrabarti, S.K. and Sheikh, A.Y., 1993,
  \art{Force on a Charged Particle Orbiting Around a Kerr-Newman 
       Black Hole}
  in \book{Proceedings of the Sixth Marcel Grossmann Meeting 
            on General Relativity}
  \Eds{H. Sato and T. Nakamura}
  \pub(World Scientific Press, Singapore), p.~1360.

\item[ ] Chakrabarti, S.K.,  1993, 
  \art{Reversal of Force and Energy Coupling 
       Around a Rotating Black Hole}
   \MNRAS \vol{261}, 625.

\item[ ] Choquet-Bruhat, Y., 1956, 
  \art{Sur L'Integration des \'Equations de la Relativit\'e G\'en\'erale}  
  \journal{J.\ Rat.\ Mech.\ Anal.} \vol{5}, 951.

\item[ ] Choquet-Bruhat, Y. and York, J.W., Jr., 1980, 
  \art{The Cauchy Problem}
  in \book{General Relativity and Gravitation} Vol. I, 
  \Ed{A. Held}
  \pub(Plenum, New York).

\item[ ] Christodoulou, D., and Ruffini, R., 1973, 
  \art{On the Electrodynamics of Collapsed Objects}
  in \book{Black Holes}, 
  \Eds{C. DeWitt and B.S. DeWitt}
  \pub(Gordon and Breach, New York), R151.

\item[ ] Ciufolini, I., 1986,
  \art{Generalized Geodesic Deviation Equation}
  \PR \vol{D34}, 1014.

\item[ ] Ciufolini, I., 1990, 
  \art{General Relativistic Measurements with Satellite Laser Ranging, 
    Lunar Laser Ranging and Very Long Baseline Interferometry} 
  \NCIM \vol{13}C, 67.

\item[ ] Ciufolini, I., 1994, 
  \art{On Gravitomagnetism and Dragging of Inertial Frames}
  \CQG \vol{11}, A73. 

\item[ ] Ciufolini, I. and Demia\'nski, M.,  1986,
  \art{How to Measure the Curvature of Space-time}
  \PR \vol{D34}, 1018. Erratum: \PR \vol{D35}, 773.

\item[ ] Ciufolini, I., Chieppa, F., Lucchesi, D. and Vespe, F., 1997,
  \art{Test of Lense-Thirring Orbital Shift Due to Spin}
  \CQG \vol{14}, 2701.

\item[ ] Ciufolini, I. and Wheeler, J.A., 1995,
   \book{Gravitation and Inertia}
   \pub(Princeton University Press, Princeton).

\item[ ] Cohen, J.M. and Toton, E.T., 1971, 
  \art{Pulsar Electrodynamics}
   \journal{Astrophys.\ Lett.} \vol{7}, 213.

\item[ ] Cohen, J.M. and Mashhoon, B., 1993,
  \art{Standard Clocks, Interferometry, and Gravitomagnetism}
  \PL \vol{A181}, 353.

\item[ ] Cohen, J.M., Tiomno, J., and Wald, R., 1973, 
  \art{Gyromagnetic Ratio of a Massive Body}
  \PR \vol{7}, 998.  

\item[ ] Cohen, J. M. and Mashhoon B., 1993
\art{Standard Clocks, Interferometry, and Gravitomagnetism}
  \PL \vol{A181}, 353.

\item[ ] Collins, C.B., and Szafron, D.A., 1979,
  \art{A New Approach to Inhomogeneous Cosmologies: 
       Intrinsic Symmetries. I.}
  \JMP \vol{20}, 2347.

\item[ ] Collins, C.B., and Szafron, D.A., 1979,
  \art{A New Approach to Inhomogeneous Cosmologies: Intrinsic Symmetries. 
    III. Conformally Flat Slices and Their Analysis}
  \JMP \vol{20}, 2347.

\item[ ] Corum, J.F., 1977, 
  \art{Relativistic Rotation and the Anholonomic Object}
  \JMP \vol{18}, 770.

\item[ ] Corum, J.F., 1980,                 
   \art{Relativistic covariance and rotational electrodynamics}
  \JMP \vol{18}, 2360.

\item [D]

\item[ ] Damour, T., 1982, 
  \art{Surface Effects in Black Hole Physics}
  in \book{Proceedings of the Second Marcel Grossmann Meeting
  on General Relativity}
  \Ed{ R. Ruffini}
  \pub(North Holland, Amsterdam),  587.

\item[ ] Damour, T., 1978, 
  \art{Black Hole Eddy Currents}  \PR \vol{D18}, 3598.

\item[ ] Damour, T., Hanni, R.H., Ruffini, R.,  and  Wilson, J.R., 1978,
  \art{Regions of Magnetic Support of a Plasma around a Black Hole}
  \PR \vol{17}, 1518.

\item[ ] Damour, T. and Ruffini, R.,  1975, 
  \art{Quantum Electrodynamical Effects in Kerr-Newmann Geometries}
  \PRL \vol {35}, 463.

\item[ ] Damour, T., 1987,
  \art{??}
  in \book{300 Years of Gravitation}
  \Eds{S.W. Hawking and W. Israel} 
  \pub(Cambridge University Press, Cambridge), 128.

\item[ ] Damour, T. Soffel, M. and Xu, C., 1991,
  \art{General Relativistic Celestial
  Mechanics I. Method and Definition of Reference Systems}
  \PR \vol{D43}, 3273.

\item[ ] Damour, T. Soffel, M. and Xu, C., 1992,
  \art{General Relativistic Celestial Mechanics II. Translational Equations
        of Motion}
  \PR \vol{D45}, 1017.

\item[ ] Damour, T. Soffel, M. and Xu, C., 1993,
  \art{General Relativistic Celestial Mechanics III. Rotational Equations
        of Motion}
  \PR \vol{D47}, 3124.

\item[ ] Damour, T. Soffel, M. and Xu, C., 1993,
  \art{New Approach to the General Relativistic $N$-Body Problem}
  in \book{Proceedings of the First William Fairbank Meeting (1990)}
  \pub(World Scientific Press, Singapore).

\item[ ] Damour, T. Soffel, M. and Xu, C., 1993,
  \art{Relativistic Celestial Mechanics}
  in \book{Proceedings of the Sixth Marcel Grossmann Meeting 
            on General Relativity}
  \Eds{H. Sato and T. Nakamura}
  \pub(World Scientific Press, Singapore), 1282.

\item[ ] de Felice, F.,  1971, 
  \art{On the Gravitational Field Acting as an Optical Medium}
  \GRG \vol{2}, 347. 

\item[ ] de Felice, F., 1975,
  \art{Analogia fra campi gravitationali e campi 
       electromagnetici}
  \journal{Rend.\ Sc.\ Fis.\ Mat.\ e Nat.\ Accad.\ 
           Naz.\ dei Lincei} \vol{58} (serie 8), 231.

\item[ ] de Felice, F.,   1979, 
  \art{On the Nonexistence of Nonequatorial Circular Geodesics with 
        Constant Latitude in the Kerr Metric}
   \PL \vol{96A}, 307.  

\item[ ] de Felice, F., 1990,
  \art{On the Circular Motion in General Relativity: Energy Threshold and
       Gravitational Strength}
  \journal{Rendiconti Matematica} Roma Serie VII \vol{10}, 59. 

\item[ ] de Felice, F., 1991,
  \art{Rotating Frames and Measurements of Forces in General Relativity}
  \MNRAS \vol{252}, 197.

\item[ ] de Felice, F., 1994,
  \art{Kerr Metric: The Permitted Angular Velocity Pattern and
     the Pre-Horizon Regime}
  \CQG \vol{11}, 1283.

\item[ ] de Felice, F., 1995,
  \art{Circular Orbits: A New Relativistic Effect in the Weak Gravitational
     field of a Rotating Source}
  \CQG \vol{12}, 1119.

\item[ ] de Felice, F.  and Bradley, M.,  1988, 
  \art{Rotational anisotropy and repulsive effects in the Kerr metric}
  \CQG \vol{5}, 1577. 

\item[ ] de Felice, F. and Calvani, M.,  1979, 
  \art{Causality violation in the Kerr metric}
  \GRG \vol{10}, 335.

\item[ ] de Felice, F., and Semar\'ak, O., 1997,
  \art{Quasi-local Measurements and Orientation in Black-hole Fields}
  \CQG \vol{14}, 2381.

\item[ ] de Felice, F., and Usseglio-Tomasset, S., 1991,
  \art{On the Pre-Horizon Regime in the Kerr Metric}
  \CQG \vol{8}, 1871.

\item[ ] de Felice, F., and Usseglio-Tomasset, S., 1993,
  \art{Schwarzchild Spacetime: Measurements in Orbiting Space-Stations}
  \CQG \vol{10}, 353.

\item[ ] de Felice, F.  and Usseglio-Tomasset, S.,  1996, 
  \art{Strains and Rigidity in Black Hole Fields}
  \GRG \vol{28}, 179. 

\item[ ] Demiansky, M., 1985,  
  \book{Relativistic Astrophysics} 
  \pub(Permagon Press, New York).

\item[ ] de Sitter,W., 1916,
 \MNRAS \vol{76}, 155, 481.

\item[ ] Deutsch, A.J., 1955, 
\art{The Electromagnetic Field of an Idealized Star in Rigid Rotation in Vacuo}
 \journal{Ann.\ Astrophys.} \vol{18}, 1.

\item[ ] Dirac, P.A.M., 1958, 
  \art{Generalized Hamiltonian Dynamics}
  \journal{Proc.\ Roy.\ Soc.\ London} \vol{A246}, 326.

\item[ ] Durrer, R. and Straumann, S., 1988, 
  \art{Some  Applications of the $3+1$
    Formalism of General Relativity}
  \journal{Helv.\ Phys.\ Acta.} \vol{61}, 1027.

\item[ ] Durrer, R., 1989, 
  \art{Gauge-Invariant Cosmological Perturbation Theory 
    for Collisionless Matter: Numerical Results} 
  \APJ \vol{208}, 1.

\item[E]
\item[ ] Ehlers, J., 1961, 
  \art{Beitr\"age zur relativistischen Mechanik kontinuierlicher Medien
        (Contributions to relativistic continuum mechanics)}
  \journal{Akad.\ Wiss.\ Mainz Abh., Math.-Nat.\ Kl.\ }  \vol{11}, 793.
  [English translation by G.F.R.\ Ellis: 1993, 
  \GRG \vol{25}, 1225.]

\item[ ] Ehlers, J., 1997, 
  \art{Examples of Newtonian Limits of Relativistic Spacetimes}
  \CQG \vol{14}, A119.

\item[ ] Ellis, G.F.R.,
  see also Bruni, Rothman.

\item[ ] Ellis, G.F.R., 1971, 
  \art{Relativistic Cosmology}
  in \book{General Relativity and Cosmology: Proceedings
    of Course 47 of the International School of Physics ``Enrico Fermi"}
  \Ed{R. Sachs}
  \pub(Academic Press, New York).

\item[ ] Ellis, G.F.R., 1973,   
  \art{Relativistic Cosmology}
  in \book{C\`argese Lectures in Physics}, Vol. 6,
  \Ed{E. Schatzman}
  \pub(Gordon and Breach, New York).

\item[ ] Ellis, G.F.R. and Bruni, M., 1989, 
  \art{A Covariant and Gauge-invariant
    Approach to Cosmological Density Fluctuations} 
  \PR \vol{D40}, 1804.

\item[ ] Ellis, G.F.R., Bruni, M., and Hwang, J., 1990, 
  \art{Density-Gradient-Vorticity Relation in Perfect Fluid
    Robertson-Walker Perturbations},
  \PR \vol{D42}, 1035.

\item[ ] Ellis, G.F.R., and Dunsby, P.K.S., 1994,
  \art{Newtonian Evolution of the Weyl Tensor}
\ASTROPH 9410001.

\item[ ] Ellis, G.F.R., Hwang, J. and Bruni, M., 1989, 
  \art{Covariant and Gauge-Independent Perfect-Fluid Robertson-Walker
    Spacetimes} 
  \PR \vol{D40}, 1819.

\item[ ] Ellis, G.F.R., and MacCallum, M.A.H., 1969,
  \art{A Class of Homogeneous Cosmological Models}
  \CMP \vol{12}, 108.

\item[ ] Ellis, G.F.R., and van Elst, H., 1998,
  \art{Cosmological Models (Carg\`{e}se lectures 1998)}
  in \book{Theoretical and Observational Cosmology}, 
  \Ed{Marc Lachi\`{e}ze-Rey}, 
  \pub(Kluwer, Dordrecht), 1.


\item[ ] Estabrook, F.B.,  and Wahlquist, H.D., 1964, 
  \art{Dyadic Analysis of Space-time Congruences}
  \JMP \vol{5}, 1629.

\item[ ] Everitt, C.W.F., 1979, 
  \art{The Gyroscope Experiment -- I: General Description and Analysis 
    of Gyroscope Performance}
  in \book{Experimental Gravitation}
  \Ed{B. Bertotti}
  \pub(Academic Press, New York).

\item[ ] Everitt, C.W.F.,  1979, 
  in \book{Experimental Gravitation: Proceedings of
Course 56 of the International School of Physics ``Enrico Fermi"}
  \Ed{B. Bertotti} 
  \pub(Academic Press, New York).

\item [F]

\item[ ] Ferraris, M., Francaviglia, M., and Sinicco, I., 1993,
  \art{Covariant ADM Formulation Applied to General Relativity}
  \NCB \vol{107}, 1303.

\item[ ] Fermi, E., 1922, 
 \art{Sopra i fenomeni che avvengono in vicinanza di una linea oraria} 
 \journal{Atti Accad.\ Naz.\ Lincei Cl.\ Sci.\ Fis.\ Mat.\ e \ Nat.} 
    \vol{31}, 184, 306.

\item[ ] Ferrarese, G., 1963, 
  \art{Contributi alla tecnica delle proiezioni in una variet\`a riemanniana 
     a metrica iperbolica normale}
  \journal{Rend.\ di Mat.} \vol{22}, 147.

\item[ ] Ferrarese, G., 1965, 
  \art{Propriet\`a di secondo ordine di un generico riferimento fisico 
     in Relativit\`a generale}
  \journal{Rend.\ di Mat.} \vol{24}, 57.

\item[ ] Ferrarese, G., 1987, 
  \art{Intrinsic Formulations in Relativistic Continuum Mechanics}   
  in \book{Selected Problems in Modern Continuum Theory} 
  \Eds{W. Kosinsky, T. Manacorda, A. Morro and T. Ruggeri}
  \pub(Pitagora Editrice, Bologna).

\item[ ] Ferrarese, G., 1988, 
  \art{Intrinsic Formulation for the Cauchy Problem in General Relativity}
  \journal{C.\ R.\ Acad.\ Sci.\ Paris (Ser I)}, \vol{307}, 107.

\item[ ] Ferrarese, G., 1989, 
  \art{Intrinsic Formulation of the Cauchy Problem in General Relativity} 
  in \book{Proceedings of the Fifth Marcel Grossmann Meeting on General 
    Relativity} 
  \Eds{D.G. Blair and M.J. Buckingham}
  \pub(World Scientific, Singapore).

\item[ ] Fischer, A.E.  and Marsden, J.E., 1978, 
  \art{Topics in the Dynamics of General Relativity}
  in  \book{Isolated Gravitating Systems in General Relativity},
  \Ed{J. Ehlers}
  \pub(Italian Physical Society, Bologna).

\item[ ] Fischer, A.E.,  and Marsden, J.E., 1979, 
  \art{The Initial Value Problem and the Dynamical Formulation of General
    Relativity}
  in \book{General Relativity: An Einstein Centenary Survey}
  \Eds{S.W. Hawking and W. Israel}
  \pub(Cambridge University Press, Cambridge).

\item[ ] Fodor, G., and Perj\'es, Z., 1994,
  \art{Canonical Gravity in the Parametric Manifold Picture}
  \GRG \vol{26}, 759.

\item[ ] Fokker, A.D., 19??,
  \journal{Proc.\ Roy.\ Acad.\ Amsterdam} \vol{23}, 379.

\item[ ] Forward, R.L., 1961, 
  \art{General Relativity for the Experimentalist}
  \journal{Proceedings of the IRE} \vol{49}, 892.

\item[G]

\item[ ] Gerosh, R., 1971, 
  \art{A Method for Generating Solutions of Einstein's Equations}  
  \JMP \vol{12}, 918.

\item[ ] G\"odel, K., 1949, 
  \art{An Example of a New Type of Cosmological Solutions of Einstein's 
     Field Equations of Gravitation} 
  \RMP \vol{21}, 447.

\item[ ] Goode, S.W., 1989, 
  \art{Analysis of Spatially Inhomogeneous Perturbations of the FRW  
     Cosmologies} 
  \JMP \vol{39}, 2882.

\item[ ] Gotay, M.J., Isenberg, J., Marsden, J.E., Montgomery, R., \'Sniatycki,
    J., and Yasskin, P.B., 1991,
  \book{Momentum Maps and Classical Relativistic Fields} (to appear).

\item[ ] Greene, R.D., Sch\"ucking, E.L. and Vishveshwara, C.V., 1975,
  \art{The Rest Frame in Stationary Space-times With Axial Symmetry}  
  \JMP \vol{16}, 153.

\item[ ] Gupta, A.,  Iyer, S.  and Prasanna A.R.,  1996, 
  \art{Centrifugal Force and Ellipticity Behaviour of a 
       Slowly Rotating Ultra Compact Object}
  \CQG \vol{13}, 2675. 

\item[ ] Gupta, A.,  Iyer, S.  and Prasanna A.R.,  1997, 
  \art{Behaviour of the Centrifugal Force and of Ellipticity 
     for a Slowly Rotating Fluid Configuration with Different 
     Equations of State}
   \CQG \vol{14}, L143.
 
\item[H]
\item[ ] Hanni, R.S., 1977, 
  \art{Wavefronts Near a Black Hole}
  \PR \vol{D16}, 933.

\item[ ] Hanni, R., and Ruffini, R., 1973, 
  \art{Lines of Force of a Point Charge Near a Schwarzschild Black Hole}
  \PR \vol{D8}, 3259.

\item[ ] Hanni, R., and Ruffini, R., 1975, 
  \art{Schwarzschild Black Hole in an Asymptotically Uniform Magnetic Field}
  \journal{Lett.\ Nuovo Cim.} \vol{15}, 189.

\item[ ] Hasse, W. and Perlick, V., 1990, 
  \art{On Redshift and Parallaxes in General Relativistic Kinematical 
    World Models} 
  \JMP \vol{31}, 1962.

\item[ ] Hawking, S.W., 1966, 
  \art{Perturbations of an Expanding Universe}
  \APJ \vol{145}, 544.

\item[ ] Hawking, S.W., and Ellis, G.F.R., 1973, 
  \book{The Large Scale Structure of Space-Time} 
  \pub(Cambridge University Press, Cambridge).

\item[ ] Henriksen, R.H., and Nelson, L.A., 1985,  
  \art{Clock Synchronization by Accelerated Observers: Metric Construction
    for Arbitrary Congruences of World Lines}  
  \CJP \vol{63}, 1393.

\item[ ] Hill, E.L., 1946, 
  \art{A Note on the Relativistic Problem of Uniform Rotation}
  \PR \vol{69}, 488.

\item[ ] Honig, E., Sch\"ucking, E.L. and Vishveshwara, C.W., 1974,
  \art{Motion of Charged Particles in Homogeneous Electromagnetic Fields}
  \JMP \vol{15}, 744.

\item[I]

\item[ ] Irvine, W.M., 1964,
  \art{Electrodynamics in a Rotating System of Reference}
  \journal{Physica} \vol{30}, 1160.

\item[ ] Isenberg, J.,  and Nester, J., 1980, 
  \art{Canonical Gravity}
  in \book{General Relativity and Gravitation} Vol. I, 
  \Ed{A. Held}
  \pub(Plenum, New York).

\item[ ] Iyer, B.R. and Vishveshwara, C.V., 1993,
  \art{Frenet-Serret Description of Gyroscopic Precession}
  \PR \vol{D48}, 5706.

\item[] Iyer, S. and Prasanna, A.P., 1993,
  \art{Centrifugal Force in Kerr Geometry}
  \CQG \vol{10}, L13.

\item[ ] Iorio, L., 2000,
\journal{Int.\ J.\ Mod.\ Phys.} \vol{D}, in press 
(\GRQC 0007014 and 0007057).

\item[J]

\item[ ] Jantzen, R.T.,
  see also Bini.

\item[ ] Jantzen, R.T., 1983, 
  \art{Perfect Fluid Sources for Spatially Homogeneous Spacetimes}  
  \AOP \vol{145}, 378.

\item[ ] Jantzen, R.T., 1990, 
  \art{Understanding Spacetime Splittings and Their Relationships}
  in \book{Fisica Matematica Classica e Relativit\'a: 
    Rapporti e Compatibilit\'a}  
  \Eds{G. Ferrarese and C. Cattani}
  \pub(Springer-Verlag, New York).

\item[ ] Jantzen, R.T., Bini, D., and Carini, P., 1996,
  \art{Gravitoelectromagnetism: Just a Big Word?}
  in \book{Proceedings of the Seventh Marcel Grossmann Meeting 
        on General Relativity (1994)}
  \Eds{R.T. Jantzen and G.M. Keiser} 
  \pub(World Scientific, Singapore), 133.

\item[ ] Jantzen, R.T. and Carini, P., 1991,
  \art{Understanding Spacetime Splittings and Their 
       Relationships}
  in \book{Classical Mechanics and Relativity: Relationship 
        and Consistency}
  \Ed{G. Ferrarese}
  \pub(Bibliopolis, Naples), 185.

\item[ ] Jantzen, R.T., Carini, P., and Bini, D., 1992,
  \art{The Many Faces of Gravitoelectromagnetism}
  \AP \vol{215}, 1.

\item[ ] Jantzen, R.T., Carini, P., and Bini, D., 1993,
  \art{Gravitoelectromagnetism: Relativity of Splitting 
       Formalisms}
  in \book{Proceedings of the Sixth Marcel Grossmann Meeting 
        on General Relativity (1991)}
  \Eds{H. Sato and T. Nakamura}
  \pub(World Scientific, Singapore), 135.

\item[ ] Jantzen, R.T., Carini, P., and Bini, D., 1993,
  \art{Gravitoelectromagnetism: Applications to Rotating 
       Minkowski, G\"odel, and Kerr Spacetimes}
  in \book{Proceedings of the Sixth Marcel Grossmann Meeting 
        on General Relativity (1991)}
  \Eds{H. Sato and T. Nakamura}
  \pub(World Scientific, Singapore), 1622.

\item[ ] Jantzen, R.T., Carini, P., and Bini, D., 2010,
  \book{Understanding Spacetime Splittings and Their 
        Relationships}
  (in preparation).

\item[K]

\item[ ] Karlovini, M.,  Rosquist, K. and Samuelsson, L., 2000,
    \art{Ultracompact stars with multiple necks}
    \journal{Ann.\ Physik} \vol{9}, 149
   (see also \GRQC 0009073 and \GRQC 0009079).

\item[ ] Kichenassamy, S. and Krikorian, R.A., 1991,
  \art{The Relativistic Rotation Transformation and the Corotating Source
  Model}
  \APJ \vol{371}, 277.

\item[ ] King, A.R. and Ellis, G.F.R., 1973, 
  \art{Tilted Homogeneous Cosmological Models}
  \CMP \vol{31}, 209.

\item [ ] Kristiansson, S., Sonego, S., Abramowicz, M.A., 1998,
  \art{Optical Space of the Reissner-Nordstr\"om Solutions}
  \GRG \vol{30}, 275.

\item[ ] Kuang, Z. and Liang, C., 1993,
  \art{All Space-times Admitting Strongly Synchronizable Reference Frames
       Are Static}
  \JMP \vol{34}, 1016.

\item[ ] Kundt, W., and  Tr\"umper, M., 1962, 
  \journal{Akad.\ Wiss.\ Mainz Abh., Math.-Nat.\ Kl.}  Nr. 12, 196.

\item[ ] Kramer D., Stephani H., Herlt E. and MacCallum M.A.H., 1980,
\book{Exact Solutions of Einstein's Theory}
  \Ed{E. Schmutzer}
  \pub(Cambridge University Press, Cambridge).

\item[L]
\item[ ] Landau, L.D., and Lifshitz, E.M., 1941, 
  \book{Teoriya Polya}
  \pub(Nauka, Moscow).

\item[ ] Landau, L.D., and Lifshitz, E.M., 1975, 
  \book{The Classical Theory of Fields} 
\pub(Permagon Press, New York).

\item[ ] Lense, J.  and Thirring, H., 1918,
\art{\"Uber den Einfluss der
  Eigenrotation der Zentralk\"orper auf die Bewegung der Planeten und Monde
  nach der Einsteinschen Gravitationstheorie} 
  \journal{Phys.\ Zeitschr.} \vol{19}, 156.

\item[ ] Lesame, W.M., 1995,
\art{Irrotational Dust with a Purely Magnetic
    Weyl Tensor}
\GRG \vol{27}, 1111; erratum, 1995,
\GRG \vol{27}, 1327.

\item[ ] Lesame, W.M., Dunsby, P.K.S., and Ellis, G.F.R, 1995,
  \art{Integrability Conditions for Irrotational Dust with a Purely Electric
       Weyl Tensor: A Tetrad Analysis}
   \PR \vol{D52}, 3406.

\item[ ] Lesame, W.M., Ellis, G.F.R, and Dunsby, P.K.S., 1996,
  \art{Irrotational Dust with ${\rm div}\, H = 0$}
   \PR \vol{D53}, 738.

\item[ ] Lesame, W.M., Ellis, G.F.R, and Dunsby, P.K.S., 1994,
  \art{Irrotational Dust with a Zero Circulation Magnetic
       Weyl Tensor}
  \pub(Dept. of Applied Mathematics, University of Cape Town).

\item[ ] 
  Lewis, T., 1932, 
  \journal{Proc.\ Roy.\ Soc.\ Lond.} \vol{331}, 176 (1932).

\item[ ] Li, N. and Torr, D.G., 1991, 
  \art{Effects of a Gravitomagnetic Field on Pure Superconductors}  
  \PR \vol{D43}, 457.

\item[ ] Lichnerowicz, A., 1944, 
  \art{L'int\'egration des \'equations de la gravitation relativiste et 
     le probl\`eme des $n$ corps}
  \journal{J.\ Math.\ Pures Appl.} \vol{23}, 37.

\item[ ] Lichnerowicz, A., 1955, 
  \book{Th\'eories Relativistes de la Gravitation et de L'Electromagn\'etisme} 
  \pub(Masson, Paris).

\item[ ] Lichnerowicz, A., 1967, 
  \book{Relativistic Hydrodynamics and Magnetohydrodynamics}
  \pub(Benjamin, New York).

\item [ ] Lichtenegger. H.I.M., Gronwald, F. and Mashhoon, B.,   2000,
  \journal{Adv.\ Space Res.} \vol{25}, 1255.

\item[ ] Lottermoser, M., 1988, 
  \art{On the Newtonian Limit of General Relativity
     and the Relativistic Extension of Newtonian Initial Data}
  Ph.D. Dissertation 
  \pub(Ludwig-Maximilians-Universit\"at, Munich).

\item[ ] Lottermoser, M., 1992,
  \art{The Post-Newtonian Approximation for the Constraint Equations 
    in General Relativity}
  \journal{Ann.\ Inst.\ H.\ Poin.} \vol{57}, 279.

\item[M]

\item[ ] Maartens, R., 1997,
  \art{Linearisation instability of gravity waves?}
  \PR \vol{D55}, 463.

\item[ ] Maartens, R. and Bassett, B.A., 1998,
  \art{Gravito-electromagnetism}
    \CQG \vol{15}, 705.

\item[ ] Maartens, R., Ellis, G.F.R., and Siklos, S.T.C., 1997,
  \art{Local Freedom in the Gravitational Field}
  \CQG \vol{14}, 1927.

\item[ ] Maartens, R., Lesame, W.M., and Ellis, G.F.R., 1997,
  \art{Consistency of Dust Solutions with div H=0}
  \PR \vol{D55}, 5219.

\item[ ] Maartens, R., Lesame, W.M., and Ellis, G.F.R., 1998,
   \art{Newtonian-like and Anti-Newtonian Universes}
   \CQG \vol{15}, 1005.

\item[ ] Maartens, R., and Triginer, J., 1997,
  \art{Density Perturbations with Relativistic Thermodynamics}
  \PR \vol{D56}, 4640.

\item[ ] MacCallum, M.A.H., 1973,
  \art{Cosmological Models From a Geometric Point of View}
  in \book{Cargese Lectures in Physics 6, Lectures at the International
    Summer School of Physics, Cargese, Corsica 1971}
  \Ed{E. Schatzman}
  \pub(New York: Gordan and Breach).

\item[ ] Macdonald, D., and Thorne, K.S., 1982,
  \art{Black-Hole Electrodynamics: an Absolute-Space/Universal-Time 
     Formulation}
 \MNRAS \vol{198}, 345.

\item[ ] Marck, J.A., 1983, 
  \art{Solution to the Equations of Parallel Transport
    in Kerr Geometry; Tidal Tensor}
  \journal{Proc.\ R.\ Soc.\ Lond.\ A} \vol{385}, 431.

\item[ ] Marck, J.A., 1983, 
  \art{Parallel-Tetrad on Null Geodesics in Kerr
       and Kerr-Newman Space-Time}
  \journal{Phys.\ Lett.} \vol{97A}, 140.

\item[ ] Marck, J.A., 1996, 
  \art{Short-cut Method of Solution of Geodesic Equations 
       for Schwarzschild black hole}
  \CQG \vol{13}, 393.

\item [ ] Martinez, E.A., 1994,
\art{Quasi-local Energy for a Kerr Black Hole}
\PR  \vol{D50}, 4920.

\item[ ] Mashhoon, B.,
  see also Cohen.

\item[ ] Mashhoon, B., 1973,
 \art{Scattering of Electromagnetic Radiation from a Black Hole}
  \PR \vol{7}, 2807.

\item[ ] Mashhoon, B., 1974,
  \art{Electromagnetic Scattering From a Black Hole and the Glory Effect}
  \PR \vol{10}, 1059.

\item[ ] Mashhoon, B., 1974,
  \art{Can Einstein's Theory of Gravitation Be Tested Beyond the
  Geometrical Optics Limit?}    
  \NAT \vol{250}, 316.

\item[ ]  Mashhoon, B., 1975,
 \art{Influence of Gravitation on the Progagation of Electromagnetic 
Radiation from a Black Hole}
  \PR \vol{11}, 2679.

\item[ ] Mashhoon, B., 1989,
  \art{Electrodynamics in a Rotating Frame of Reference}
  \PL \vol{A139}, 103.

\item[ ] Mashhoon, B.,  1992, 
  \art{On the Strength of a Gravitational Field}
  \PL \vol{A163}, 7.

\item[ ] Mashhoon, B., 1999,
  \art{On the spin-rotation-gravity coupling}
  in \book{Mexican Meeting on Gauge Theories of Gravity (Mexico City, 1997)}
 \GRG \vol{31}, 681.

\item[ ] Mashhoon, B., Gronwald, F. and Lichtenegger, H.I.M, 2000,
  \art{Gravitomagnetism and the Clock Effect}
  in \book{Gyros, Clocks and Interferometers: Testing General Relativity
  in Space},
  \Eds{C. L\"ammerzahl C.W.F. Everitt and F.W. Heyl}
  \pub(Springer, Berlin).

\item [ ] Mashhoon, B., Gronwald, F. and Theiss, D.S., 1999,
\art{On Measuring Gravitomagnetism via Spaceborne Clocks: A Gravitomagnetic
Clock Effect}
  \journal{Ann.\ Physik} \vol{8}, 135.

\item[ ] Mashhoon, B., Heyl, F.W., and Theiss, D.S., 1984,
  \art{On the Gravitational Effects of Rotating Masses: The Thirring-Lense
       Papers}
  \GRG \vol{16}, 711.

\item[ ] Mashhoon, B.,  McClune, J.C., Quevedo, H., 1997,
  \art{Gravitational Superenergy Tensor}
  \PL \vol{A231}, 47.

\item[ ] Mashhoon, B.,  McClune, J.C., Quevedo, H., 1999, 
  \art{On the Gravitoelectromagnetic Stress-energy Tensor} 
  \CQG \vol{16}, 1137.

\item[ ] Mashhoon, B., Paik, H.J. and Will, C.M., 1989,
  \art{Detection of the Gravitomagnetic Field Using an Orbiting         Superconducting Gravity Gradiometer: Theoretical Principles} 
  \PR \vol{D39}, 2825.

\item[] Mashhoon, B. and Santos, N.O., 2000
\art{Rotating Cylindrical Systems and Gravomagnetism}
  \journal{Ann.\ Physik} \vol{9}, 49.

\item[ ] Massa, E., 1974, 
  \art{Space Tensors in General Relativity I: Spatial Tensor Algebra 
     and Analysis} 
  \GRG \vol{5}, 555.

\item[ ] Massa, E., 1974, 
  \art{Space Tensors in General Relativity II: Physical Applications}   
  \GRG \vol{5}, 573.

\item[ ] Massa, E., 1974, 
  \art{Space Tensors in General Relativity III: The Structural Equations}   
  \GRG \vol{5}, 715.

\item[ ] Massa, E. and Zordan, C., 1975, 
  \art{Relative Kinematics in General Relativity: The Thomas and Fokker 
    Precessions}
  \journal{Meccanica} \vol{10}, 27.

\item[ ] Massa, E., 1990, 
  \art{Spatial Tensor Analysis in General Relativity}
  in \book{Fisica Matematica Classica e Relativit\'a: 
    Rapporti e Compatibilit\'a}  
  \Eds{G. Ferrarese and C. Cattani}
  \pub(Springer-Verlag, New York).

\item[ ] McIntosh, C.B.G, Arianrhod, R., Wade, S.T., and Hoenselaers, C., 1994,
  \art{Electric and Magnetic Weyl Tensors: Classification and Analysis}
  \CQG \vol{11}, 1555.

\item[ ] Misner, C.W., and Wheeler, J.A., 1957, 
  \art{Classical Physics as Geometry}  
  \AOP \vol2, 525.

\item[ ] Misner, C.W., Thorne, K.S. and Wheeler, J.A., 1973,
  \book{Gravitation} 
  \pub(Freeman, San Francisco).

\item[ ] Mitskievich,  N.V.,
  \art{Relativistic Physics in Arbitrary Reference Frames}
 \GRQC 9606051.

\item[ ] Mitskievich,  N.V. and  Pulido Garcia, I., 1970,
  \art{??}
  \journal{Sov.\ Phys.\ Dokl.} 15, 591. 

\item [ ]
Mitskievich, N. V. and Zaharow, V. N., 1970,
 \art{??} 
 \journal{Doklady Akad.\ Nauk.\ SSSR.}, \vol{195}, 321 (in Russian).

\item[ ] M\o ller, C., 1952,
  \book{The Theory of Relativity}, First Edition;
1972, Second Edition,
  \pub(Oxford University Press, Oxford).

\item[N]

\item[ ] Nayak, K.R. and Vishveshwara, C.V., 1996,
  \art{Gyroscopic Precession and Inertial Forces
       in the Kerr-Newman Spacetime}
  \CQG \vol{13}, 1783.

\item[ ] Nayak, K.R.  and Vishveshwara, C.V.,  1997, 
  \art{Gyroscopic Precession and Centrifugal Force
       in the Ernst Spacetime}
  \GRG \vol{29}, 291. 

\item[ ] Nordtvedt, K., 1988, 
  \art{Gravitomagnetic Interaction and Laser Ranging to Earth Satellites}
  \PRL \vol{61}, 2647.

\item[ ] Nordtvedt, K., 1988, 
  \art{Existence of the Gravitomagnetic Interaction}
  \IJTP \vol{27}, 1395 (1988).

\item[O]

\item[ ] Olson, D.W., 1976, 
  \art{Density Perturbations in Cosmological Models}
  \PR \vol{D14}, 327.

\item[ ] Ozsv\'ath, I., 1977, 
  \art{Spatially Homogeneous Lichnerowicz Universes}
  \GRG \vol{8}, 737.

\item[ ] Ozsv\'ath, I.,  and Sch\"ucking, E., 1969,  
  \art{The Finite Rotating Universe}   
  \AOP \vol{55}, 166. 

\item[P]

\item[ ] Page, D.N.,  1993, 
  \art{??}
  \journal{Sci.\ Am.} \vol{269}, 10. 

\item[ ]
Page, D., 1998,
  \art{Maximal acceleration is nonrotating}
\CQG \vol{15}, 1669. 

\item[ ] Papapetrou, A., 1966, 
  \journal{Ann.\ Inst.\ H.\ Poincar\'e} \vol{A4}, 83.

\item[ ] Perj\'es, Z., 1989, \art{Parametric Manifolds}
  in \book{Proceedings of the Fifth Marcel Grossmann Meeting on 
     General Relativity}
  \Eds{D.G. Blair and M.J. Buckingham}
  \pub(World Scientific, Singapore).

\item[ ] Perj\'es, Z., 1993, 
  \art{The Parametric Manifold Picture of Space-Time}
  \NUCP \vol{B403}, 809. !-837

\item[ ] Perlick, V., 1990, 
  \art{On Fermat's Principle in General Relativity: I. The General Case}
  \CQG \vol{7}, 1319.

\item[ ] Perlick, V., 1990, 
  \art{On Fermat's Principle in General Relativity: II. The Conformally 
     Stationary Case}
  \CQG \vol{7}, 1849.

\item[ ] Perlick, V., 1990, 
  \art{Geometrical and Kinematical Characterization of Parallax-Free World 
     Models}
  \JMP \vol{29}, 2064.

\item[ ] Perlick, V. 1991, 
  \art{A Class of Stationary Charged Dust Solutions of Einstein's 
       Field Equations}
  \GRG \vol{23}, 1337.

\item[ ] Pietronero, L., 1973,
  \art{The Mechanics of Particles Inside a Rotating Cylindrical Mass Shell}
  \AOP \vol{79}, 250.

\item[ ] Pietronero, L., 1974,
  \art{Mach's Principle for Rotation}
  \NCIM \vol{20B}, 144.

\item[ ] Pirani, F.A.E., 1956,
  \journal{Acta Physica Polonica} \vol{15}, 389.

\item[ ] Plebansky, J., 1960, 
  \art{Electromagnetic Waves in Gravitational Fields}
  \PR \vol{118}, 1396.

\item[ ] Post, E.J., 1967, 
  \art{Sagnac Effect} 
  \RMP \vol{39}, 475.

\item[ ] Prasanna, A.R.,
  see also Iyer.

\item[ ] Prasanna, A.R.,  1991, 
  \art{Centrifugal Force in Ernst Space-Time}
  \PR \vol{D43}, 1418. 

\item[ ] Prasanna, A.R., 1997,
  \art{Inertial Frame Dragging and Mach's Principle in General
       Relativity}
  \CQG \vol{14}, 227.

\item[ ] Prasanna, A.R. and Chakrabarti, S.K., 1990,
  \art{Angular Momentum Coupling and Optical Reference Geometry
    in Kerr Spacetime}  
  \GRG \vol{22}, 987.
   
\item[ ] Prasanna, A.R.  and Iyer, S.,  1991, 
   \art{The Radial Force on a Charged Particle
     in Superimposed Magnetic Fields on Schwarzschild Space-Time}
   \journal{Pramana J.\ Phys} \vol{37}, 405. 

\item[ ] Prasanna, A.R. and Iyer, S., 1997,
  \art{Cumulative Dragging: An Intrinsic Characterization of
       Stationary Axisymmetric Spacetime}
  \PL \vol{A233}, 17.

\item[Q]

\item[ ] Qadir, A., 1989,
  \book{Introduction to Special Relativity}
  \pub(World Scientific, Singapore).

\item[R]
\item[ ] Raychaudhuri, R., 1955, 
  \PR \vol{98}, 1113.

\item[ ] Raychaudhuri, R., 1957, 
  \journal{Zeits.\ f.\ Astrophys.} \vol{48}, 161.

\item[ ] Rindler, W., 1977,
  \art{The Case Against Space Dragging}
  \PL \vol{A233}, 25.

\item[ ] Rindler, W., 1977,
  \book{Essential Relativity} Second Revised Edition,
  \pub(Springer-Verlag, New York).

\item[ ] Rindler, W. and Perlick, V., 1990,
  \art{Rotating Coordinates as Tools for Calculating Circular Geodesics
     and Gyroscopic Precession}
  \GRG \vol{22}, 1067.

\item [ ] Romano, J.D., Price, R.H., 1995,
  \art{Embedding Initial Data for Black-Hole Collisions}
  \CQG \vol{12}, 875.

\item[ ] Rothman, T., Ellis, G.R.F. and Murugan, J., 2000,
  \art{Holonomy in the Schwarzschild-Droste Geometry}
  \GRQC 0008070.

\item[ ] Ruffini, R.,  and Wilson, J.R., 1975, 
  \art{Relativistic Magnetohydrodynamical Effects of Plasma 
     Accreting into a Black Hole}
  \PR \vol{D12}, 2959.

\item[ ] Ruffini, R., 1978, 
  in \book{Physics and Astrophysics of Neutron Stars and Black Holes}, 
  \Eds{R. Giacconi and R. Ruffini}
  \pub(Italian Physical Society, Bologna).

\item[ ] Ryan Jr, M.P., and Shepley, L.C., 1975, 
  \book{Homogeneous Relativistic Cosmologies} 
  \pub(Princeton University Press, Princeton).

\item[S]

\item[ ] Sachs, R.K., and Wu, H., 1977, 
  \book{General Relativity for Mathematicians} 
  \pub(Springer-Verlag, New York).

\item[ ] Samuel, J. and Iyer, B.R., 1986, 
  \art{A Gravitational Analog of the Dirac Monopole} 
  \journal{Current Science} \vol{55}, 818.

\item[ ] Schiff, L.I., 1967,
  \art{Motion of a Gyroscope According to Einstein's Theory of Gravitation}
  \journal{Proc.\ Nat.\ Acad.\ Sci.\ Am.} \vol{46}, 871.

\item[ ] Schmutzer, E., 1968, 
  \art{New Approach to Interpretation Problems of
    General Relativity by Means of the Splitting-up-formalism of Space-time}
  in \book{Induction, Physics and Ethics}
  \Eds{P. Weingartner and G. Zecha}
  \pub(Riedel, Dordrecht).

\item[ ] Schmutzer, E. and Plebansky, J., 1977, 
  \art{Quantum Mechanics in Non-inertial Frames of Reference}
  \journal{Fortschritte der Physik} \vol{25}, 37.

\item[ ] Schouten, J.A., 1921,
  \art{??}
  \journal{Math.\ Z.} \vol{11}, 55.

\item[ ] Semer\'ak, O.,
  see also de Felice.

\item[ ] Semer\'ak, O., 1993,
  \art{Stationary Frames in the Kerr Field}
  \GRG \vol{25}, 1041.

\item[ ] Semer\'ak O  1994 
   \art{On the Competition of Forces in the Kerr Field}
   \AAA \vol{291}, 679.
  
\item[ ] Semer\'ak, O.,  1995, 
   \art{On the Occurrence of Rotospheres in the Kerr field}
   \journal{Physica Scripta} \vol{52}, 488.  

\item[ ] Semer\'ak, O., 1995,
  \art{What Forces Drive Relativistic Motion?}
  \NCB \vol{110}, 973.

\item[ ] Semer\'ak, O., 1996,
  \art{Extremally Accelerated Observers in Stationary
       Axisymmetric Spacetimes}
  \GRG \vol{28}, 1151.

\item[ ] Semer\'ak, O., 1996,
  \art{What Forces Act in Relativistic Gyroscope Motion?}
  \CQG \vol{13}, 2987.

 \item[ ] Semer\'ak, O.,  1996, 
   \art{Collimation (and Other) Effects of the Kerr Field:
        an Interpretation}
   \journal{Astrophys. \ Lett. \ Commun.} \vol{33}, 275. 

\item[ ] Semer\'ak, O., 1997,
  \art{Gyroscope in Polar Orbit in the Kerr Field}
  \GRG \vol{29}, 153.

 \item[ ] Semer\'ak, O.,  1998, 
   \art{Rotospheres in Stationary Axisymmetric Spacetimes}
   \AP \vol{263}, 133.

\item[ ] Semer\'ak, O., 1999,
  \art{The Gravitomagnetic Clock Effect and Extremely Accelerated Observers}
  \CQG \vol{16}, 3769.

\item[ ] Semer\'ak, O.  and Bi\v c\'ak, J.,  1997, 
   \art{The Interplay Between Forces in the Kerr-Newman Field}
   \CQG \vol{14}, 3135.

\item[ ] Semer\'ak, O.  and de Felice, F.,  1997, 
   \art{Quasi-Local Measurements and Orientation in Black Hole Fields}
   \CQG \vol{14}, 2381.

\item[ ] Shahid-Saless, B., 1988,
  \art{Relativistic Effects in Local Inertial Frames Including 
       Parametrized-Post-Newtonian Effects}
  \PR \vol{D38}, 1645.

\item[ ] Shahid-Saless, B., 1990,
  \art{Local Gravitomagnetism}
  \GRG \vol{22}, 1147.

\item[ ] Shahid-Saless, B., 1992,
  \art{Local Gravitomagnetic Perturbations of the Lunar Orbit}
  \PR  \vol{D46}, 5404.

\item [] Sharp, N.A., 1981,
\art{On Embeddings of the Kerr Geometry}
 \journal{Can.\ J.\ Phys.} \vol{59}, 688.

\item[ ] Skrotsky, G.V., 1957,  
  \art{The Influence of Gravitation on the Propagation of Light}
  \journal{Sov.\ Phys.\ Dok.} \vol{2}, 226.

\item[ ] Smarr, L., and York Jr,  J.W., 1978b, 
  \art{Kinematical Conditions in the Construction of Spacetime}  
  \PR \vol{D17}, 2529.

\item[ ] Soffel, M.H., 1989, 
  \book{Relativity, Astrometry, Celestial Mechanics and Geodesy} 
  \pub(Springer-Verlag, Berlin).

\item[ ] Sonego, S.,
  see also Abramowicz, Kristiansson.

\item[ ] Sonego, S. and Abramowicz, M.A., 1998,
   \art{Maxwell Equations and the Optical Geometry}
   \JMP \vol{39}, 3158.

\item[ ] Sonego, S. and Lanza, A., 1996,
   \art{Relativistic Perihelion Advance as a Centrifugal Force}
   \MNRAS \vol{279}, L65.

\item[ ] Sonego, S. and Massar, M., 1996,
  \art{Covariant Definition of Inertial Forces: Newtonian Limit
       and Time-Dependent Gravitational Fields}
  \CQG \vol{13}, 139.

\item[ ] Sopuerta, C., Roy Maartens, R., Ellis, G. and Lesame, W., 1999,
    \art{Nonperturbative Gravito-Magnetic Fields}
    \PR \vol{D60}, 024006.

\item[ ] Stachel, J., 1980,
  \art{The Anholonomic Cauchy Problem in General Relativity}
  \JMP \vol{21}, 1776.

\item[ ] Stedman G. E., 1985,
\art{Ring Interferometric Tests of Classical and Quantum Gravity}
  \journal{Contemp.\ Phys.} \vol{26}, 311.

\item[ ] Stedman G. E., 1997,
\art{Ring Laser Tests of Fundamental Physics and Geophysics}
  \journal{Rep.\ Prog.\ Phys.} \vol{60}, 615.

\item[ ] Szafron, D.A., and Collins, C.B., 1979,
  \art{A New Approach to Inhomogeneous Cosmologies: Instrinsic Symmetries. 
    II. Conformally Flat Slices and an Invariant Classification}
  \JMP \vol{20}, 2347.

\item[T]

\item[ ] Takeno, H., 1952,
  \art{On Relativistic Theory of Rotating Disk}
  \PTP \vol{7}, 367.

\item[ ] Tamm, I., 1924, 
  \journal{J.\ Russ.\ Phys.-Chem.\ Soc.} \vol{56}, 284.

\item[ ] Tartaglia, A., 2000,
 \art{Detection of the Gravitomagnetic Clock Effect}
 \CQG \vol{17}, 783.

\item[ ] Tartaglia, A., 2000,
 \art{Influence of the Angular Momentum of Astrophysical Objects on
   Light and Clocks and Related Measurements} 
  \CQG \vol{17}, 2381.

\item[ ] Taub, A.H., 1969, 
  \art{Stability of Fluid Motions and Variational Principles}
  \book{Proceedings of the 1967 Colloque on ``Fluids 
    et champ gravitationel en relativit\'e g\'en\'erale"} No.~170 
  \pub(Cente National de la Recherche Scientifique, Paris), 57.

\item[ ] Taub, A.H., and MacCallum, M.A.H., 1972, 
  \art{Variational Principles and Spatially-Homogeneous Universes, 
    Including Rotation}
  \CMP \vol{25}, 173.

\item[ ] Thirring, H., and Lense, J., 1918, 
  \art{\"Uber den Einfluss der Eigenrotation der Zentralk\"orper auf die 
    Bewegung der Planeten und Monde nach der Einsteinschen Gravitationstheorie} 
  \journal{Phys.\ Z.} \vol{19}, 156.

\item[ ] Thomas, L.W., 1927, 
  \art{The Kinematics of an Electron With an Axis}
  \journal{Phyl.\ Mag.} \vol{3}, 1.

\item[ ] Thorne, K.S., 1981,
  \art{Experimental Gravity, Gravitational Waves, and Quantum Nondemolition:
  an Introduction}
  in \book{Quantum Optics, Experimental Gravitation, and Measurement Theory}
  \Eds{P. Meystre and M.O. Scully}
  \pub(Plenum Press, New York and London).

\item[ ] Thorne, K.S., 1989,
  \art{Gravitomagnetism, Jets in Quasars, and the Stanford
     Gyroscope Experiment} 
  in \book{Near Zero: New Frontiers in Physics}
  \Eds{J.D. Fairbank, B.S. Deaver, Jr., C.W.F. Everitt, P.F. Michelson}
  \pub(Freeman, New York).

\item[ ] Thorne, K.S., and Macdonald, D., 1982,  
  \art{Electrodynamics in Curved Spacetime: $3+1$ formulation}
  \MNRAS \vol{198}, 339.

\item[ ] Thorne, K.S., Price, R.H., and Macdonald, D.A., 1986, 
  \book{Black Holes: The Membrane Paradigm} 
  \pub(Yale University Press, New Haven).

\item[ ] Thorne, K.S., 1988, 
  in \book{Near Zero: New Frontiers of Physics} 
  \Eds{J.D. Fairbank, B.S. Deaver, Jr., C.W. Everitt, and P.F. Michelson}
  \pub(Freeman, New York).

\item[ ] Tiomno, J., 1973, 
  \art{Electromagnetic Field of Rotating Charged Bodies}
  \PR \vol{7}, 992.

\item[ ] Trautman, A., 1965, 
  \art{Foundations and Current Problems of General Relativity}  
  in \book{Brandeis Lectures on General Relativity}
  \Eds{A. Trautman, F.A.E. Pirani, H. Bondi}
  \pub(Prentice Hall, Englewood Cliffs, NJ).

\item[ ] Trocheris, M.G., 1949,
  \art{Electrodynamics in a Rotating Frame of Reference}
  \journal{Phil. Mag.} \vol{40}, 1143.

\item[ ] Tr\"umper, M., 1965,
  \art{On a Special Class of Type-I Gravitational Fields}
  \JMP \vol{6}, 584.

\item[ ] Tr\"umper, M., 1967, 
  \art{Bemerkungen \"uber scherungsfreie Str\"omungen gravitierender Gase}
  \journal{Zeits.\ f.\ Astrophys.} \vol{66}, 215.

\item[ ] Tsoubelis, D., Economou, A., and Stoghianidis, E., 1987,
  \art{Local and Global Gravitomagnetic Effects in Kerr Spacetime}
  \PR \vol{D36}, 1045.

\item[ ] Tsoubelis, D., Economou, A., and Stoghianidis, E., 1986,
  \art{The Geodetic Effect Along Polar Orbits in the Kerr Spacetime}
  \PL \vol{118}, 113.

\item[V]
\item[ ] Van Bladel, J., 1984, 
  \book{Relativity and Engineering} 
  \pub(Springer-Verlag, Berlin).

\item[ ] van Elst, H. and Uggla, C., 1997,
  \art{General Relativistic 1+3 Orthonormal Frame Approach Revisited}
   \CQG \vol{14}, 2673.

\item[ ] Vishveshwara, C.V., see also Iyer, Nayak.

\item[W]
\item[ ] Wainwright, J., 1979,
  \art{A Classification Scheme for Non-Rotating Inhomogenous Cosmologies}
  \JPAMG \vol{12}, 2015.

\item[ ] Wald, R.M., 1984
\book{General Relativity}, 
  \pub(University of Chicago, Chicago).

\item[ ] Walker, A.G., 1932, 
  \art{Relative Coordinates} 
  \journal{Proc.\ Roy.\ Soc.\ Edinburgh} \vol{52}, 345.

\item[ ] Weinberg, S., 1972, 
  \book{Gravitation and Cosmology}
  \pub(Wiley, New York).

\item[ ] J.A. Wheeler, 1964,
  in \book{Relativity, Groups, and Topology} 
  \Eds{C. DeWitt and B.S. DeWitt}
  \pub(Gordon and Breach, New York).

\item[ ] Wheeler, J.A., 1988, 
  \art{Geometrodynamic Steering Principle Reveals the Determiners of Inertia}
  \journal{Int.\ J.\ Mod.\ Phys.} A\vol{3}, 2207.

\item[ ] Weyl, H., 1917, 
\journal{Ann.\ Phys., Lpz.} \vol{54}, 117.

\item[ ] Wilkins, D. and Jacobs, M.W., 1992,
  \art{G\"odel's Gravitomagnet}
  \PR \vol{D46}, 3395.

\item[] Xu, D.Y. and Qin, Z.Y., 1998,
 \journal{Int.\ J.\ Theor.\ Phys.} \vol{37}, 1159. 

\item[Y]

\item[ ] York Jr., J.W., 1971,
  \art{Gravitational Degrees of Freedom and the Initial-Value Problem}
  \PRL \vol{26}, 1656. 

\item[ ] York Jr., J.W., 1972,
  \art{Role of Conformal Three-Geometry in the Dynamics of Gravitation}
  \PRL \vol{28}, 1082. 

\item[ ] York Jr., J.W., 1979, 
  \art{Kinematics and Dynamics of General Relativity}
  in \book{Sources of Gravitational Radiation}
  \Ed{L. Smarr}
  \pub(Cambridge University Press, Cambridge).

\item[Z]

\item[ ] Zel'manov, A.L., 1956, 
  \art{Chronometric Invariants and Frames of Reference in the General 
     Theory of Relativity}
  \journal{Dokl.\ Akad.\ Nauk.\ USSR} \vol{107}, 805
  [Eng. trans. in \journal{Sov.\ Phys.\ Doklady} \vol{1}, 227 (1956)].

\item[ ] Zel'manov, A.L., 1959, 
  in \book{Turdy Shestovo Soveshchaniya po Voprosam Kosmogonii} 
  \pub(Acad.\ Pub.\ USSR, Moscow).

\item[ ] Zel'manov, A.L., 1973, 
  \art{Kinematic Invariants and Their Relation to Chronometric Invariants in
    Einstein's Theory}  
  \journal{Sov.\ Phys.\ Dokl.} \vol{18}, 231.

\item[ ] Zhang, X.H., 1989, 
  \art{$3+1$ Formulation of General-Relativistic Perfect Magnetohydrodynamics} 
  \PR \vol{39}, 2933.

\item[ ] Znajek, R.L., 1977,
  \art{Black Hole Electrodynamics and the Carter Tetrad}
  \MNRAS \vol{175}, 457.

\end{enumerate}
\end{document}